\documentclass[sigconf,nonacm]{acmart}

\usepackage{hyperref}
\usepackage{times}  
\usepackage{helvet}  
\usepackage{courier}  
\usepackage{graphicx} 
\usepackage{natbib}  
\usepackage{caption} 
\usepackage{xcolor}
\usepackage{svg}
\usepackage{siunitx}
\usepackage{booktabs}
\usepackage{makecell}
\usepackage{tikz}

\usepackage{subcaption} 
\newcommand{\rev}[1]{\textcolor{black}{#1}} 

\begin{document}

\title{Too Little, Too Late: Moderation of Misinformation around the Russo-Ukrainian Conflict}

\author{Gautam Kishore Shahi}
\orcid{1234-5678-9012}
\affiliation{%
  \institution{University of Duisburg-Essen}
  \city{Duisburg}
  \country{Germany}
}
\email{gautam.shahi@uni-due.de}

\author{Yelena Mejova}
\orcid{0000-0001-5560-4109}
\affiliation{%
  \institution{ISI Foundation}
  \city{Turin}
  \country{Italy}}
\email{yelenamejova@acm.org}

\renewcommand{\shortauthors}{Shahi \& Mejova}


\begin{abstract}
In this study, we examine the role of Twitter as a first line of defense against misinformation by tracking the public engagement with, and the platform's response to, 500 tweets concerning the Russo-Ukrainian conflict which were identified as misinformation. 
Using a real-time sample of \num{543475} of their retweets, we find that users who geolocate themselves in the U.S.~both produce and consume the largest portion of misinformation, however accounts claiming to be in Ukraine are the second largest source.
At the time of writing, 84\% of these tweets were still available on the platform, especially those having an anti-Russia narrative.
For those that did receive some sanctions, the retweeting rate has already stabilized, pointing to ineffectiveness of the measures to stem their spread. 
These findings point to the need for a change in the existing anti-misinformation system ecosystem. We propose several design and research guidelines for its possible improvement.
\end{abstract}







\settopmatter{printacmref=false}
\maketitle



\section{Introduction}
\label{sec:1}

Wartime comes with many uncertainties---about the unfolding situation, and the capacity, intentions, and plans of all parties---as has long been acknowledged by the military strategists \cite{von1883kriege}. 
In the information age, this ``fog of war'' extends to the mass and social media, wherein propaganda, misinformation, or simply editorial bias obscure the full story. 
The alternative stories and narratives develop and spread at an unprecedented velocity, and, with the automated translation technologies, they are able to reach global audiences.

The Russian invasion of Ukraine on February 24th, 2022 has inflamed a simmering conflict between two nations, and heated up the media coverage and social media discussion on the topic.
Some messaging has been attributed to the Russian government and their communications efforts \cite{linvill2020troll,boyte2017analysis}, such as a purported existence of U.S. biolabs and the development of biological weapons in Ukraine \cite{alieva2022investigating}. 
However, the conflict has encouraged a slew of domestic political banter \cite{de2023twitter}, as well as scams \cite{crawford2022ukraine}, many of which propagated on Twitter.\footnote{\url{https://en.wikipedia.org/wiki/Disinformation_in_the_Russian_invasion_of_Ukraine}}\footnote{Twitter has been renamed as X in July 2023, but for the consistency with the literature we continue to refer to the platform as Twitter and its posts as tweets.}
\rev{Although several datasets have been compiled on misinformation during the Russo-Ukrainian conflict,\footnote{See https://conflictmisinfo.org/datasets/ for a selection of such resources} the full picture of the debunking efforts and especially information on the platform's subsequent intervention is still incomplete.
Further, earlier works have used automated tools to match tweets with claims \cite{la2023retrieving}, gazetteers of low-credibility news sources \cite{pierri2023propaganda}, or manual examination of the top popular users \cite{lai2024multilingual}, possibly introducing errors and biases.}


In this study, we contribute a dataset of \num{500} tweets identified as misinformation by some of the most prominent Western fact-checking organizations, and a record of their spread via a sample of \num{543475} of their retweets.
By enriching this data with geolocation information 
and manual annotation of their narratives, we provide a detailed view of the spread of this content in the year following the invasion. 
Crucially, we examine the actions taken (and usually not taken) by the Twitter platform, their timeliness and potential impact.
Specifically, we aim to address the following research questions:

\begin{itemize}
 \item \textbf{RQ1}. What are the attributes of the misinformation posters that may relate to their content's spread?
 \item \textbf{RQ2}. Where are the ostensible geographic origins of misinformation, and where is its audience?
 \item \textbf{RQ3}. What is the narrative directionality of this misinformation, and how does it intersect with its spread?
 \item \textbf{RQ4}. What is the response of the platform to this misinformation, and how effective is it?
\end{itemize}

We find that, among the 18 languages in the dataset, English remains the most prominent, and the U.S. remains the most common location in which the posters (and reposters) of misinformation self-identify themselves.
However, the second most active location is Ukraine, followed by Great Britain, and only then Russia. 
These tweets received \num{877579} retweets within our study window, likely achieving an audience into the millions, with a quarter of the misinformation circulating on the platform for more than 3 months since the posting of the original.
We show that the popularity of the tweets is mostly associated with the number of followers the original poster has, not whether the account is verified, whether it was created before or after the invasion, or even whether it was eventually sanctioned by the platform. 
Interestingly, the tweets posted by users claiming to be in Ukraine received more retweets than those from elsewhere, similarly those having pro-Ukrainian and anti-Russia narratives also were much more popular.
Crucially, the misinformation tweets over a year after their publication, we find that the vast majority -- 84\% -- of them are still available on the platform and that only 26 accounts (out of 428) were suspended.
Examining the temporal trends of the retweeting rates for the content that did receive some moderation, in most cases the interest already wanes by the time the post is taken down. 

Thus, our study illustrates the reach of misinformation during the first year of the invasion and the ineffectiveness of the platform in its handling. 
As such, it provides valuable evidence of the challenges socio-technical systems encounter when encountering unverified, potentially distorted information during conflicts. 
We provide the annotated list of misinformation tweets and the IDs of their retweets to the research community.\footnote{\url{https://github.com/Gautamshahi/RussoUkraineMisinfoTweets2022}}




\section{Related Work}




Social media has exacerbated the ``fog of war'' surrounding modern-day conflicts.
Since Russia's annexation of Crimea in 2014, and especially since its invasion of Ukraine's territory in 2022, social media has been a parallel battleground for the public discourse around the conflict \cite{freeman2023seeing}. 
Immediately after the invasion, communications and media researchers began compiling datasets capturing the media coverage and social media posts around the conflict.
\citet{khairova2023first} created a dataset of relevant news articles from media outlets in Ukraine, Russia, Europe, Asia, and the US around nine events. 
Similarly, there exist datasets from Twitter \cite{chen2023tweets,pohl2023invasion}, TikTok \cite{primig2023remixing}, Reddit \cite{albota2022war}, Telegram \cite{ronzhyn2023collective} and Weibo \cite{hanley2023special}.
The first year after the invasion is also the last year of Twitter API's broad availability,\footnote{\url{https://9to5mac.com/2023/04/06/twitter-shuts-down-free-api/}} allowing for potentially some of the last large-scale analyses of public discourse on the platform around the conflict.


Within this discourse, propaganda and misinformation pose a serious danger to the health of the public sphere. 
To track these, researchers employed a variety of fully and partially automated approaches. 
Researchers in the misinformation space have used manually annotated tweets about the war in the design and evaluation of ML classifiers \citet{darwish2023identifying,toraman2024mide22}.
\citet{la2023retrieving} used a strategy of retrieving tweets based on \emph{claims}, some of which may match debunked misinformation, to compile a manually-cleaned list of \num{83} claims matching \num{2359} tweets concerning the Russo-Ukrainian conflict.
However, as the authors point out, tracking claims instead of particular posts introduces a difficulty, as some claims may become true or false over time (for instance, whether NATO members supplied Ukraine with military weapons). 
Instead, \citet{lai2024multilingual} create a Twitter dataset about the war in English, Japanese, Spanish, French, German, and Korean languages 
and group tweets into clusters 
and manually assign misinformation labels to clusters which contain a tweet that can be linked to a fact-checked article that debunks it. 
A study on misinformation on Facebook and Twitter was performed using keyword-based search and annotation based on source credibility of news website mentioned in tweets \cite{pierri2023propaganda}. 
Such an approach can be considered ``distant supervision'', as the veracity of the tweet itself is not annotated.
Network properties of the datasets have also been utilized: \citet{lai2024multilingual} used the retweet network for tagging users who share misinformation URLs in their tweets, and performed explanatory analysis of tweets. 
Even before the invasion, some of the misinformation has been attributed to the Russian state actors such as Russia's Internet Research Agency \cite{linvill2020troll,boyte2017analysis} and Russian Ministry of Defense \cite{alieva2022investigating}, but a variety of topics spanning global and local politics and many kinds of sources of inaccuracy are possible \cite{lai2024multilingual}. 
During the conflict, it was found that several Twitter accounts related to Russian-based media were spreading misinformation: \citet{aguerri2024fight} filtered 90 Twitter accounts of media outlets and individual journalists and analyzed possible misinformation therein. 


A summary of Twitter misinformation datasets concerning the war can be found in Table \ref{ex:misinfodata}. 
Overall, the key challenge with source-based annotation (distant supervision) is that not all tweets are misinformation. 
For manual annotation, none of them provided the misinformation data by individually debunking tweets with background truth. 
In this study, we track a large set of misinformation tweets identified explicitly by various fact-checkers to examine their reach and the platform's response systematically. This approach allows us to examine data annotated by experts with requisite background knowledge. 
Although it limits the coverage of all potential instances of the misinformation content, it minimizes the bias in the identification of misinformation and provides clarity in the information available to the platform about specific posts. 

\begin{table*}[!htbp]
    \centering
    \small
    \caption{An Overview of Textual Datasets on misinformation on Twitter.} 
        \renewcommand{\familydefault}{\ttdefault}
\begin{tabular}{ p{2.2cm}rlp{5cm}  }  
 \toprule

Dataset & Data Collection Approach & Timeline & Annotation Technique\\ \midrule
Pierri et al. \cite{pierri2023propaganda} & Keyword based & 01/01/2022-24/04/2024 & Source Credibility \\ 

Lai et al. \cite{lai2024multilingual} & Keywords based & 24/02/2022-12/03/2024 & User who shares misinformation URLs \\

Alieva et al. \cite{alieva2022investigating}& Keyword based & 24/02/2022-08/08/2022 & Open qualitative analysis \\

La Gatta et al. \cite{la2023retrieving} & Claim based filtering & 24/02/2022-08/03/2024 & Text (claim) similarity\\

Aguerri et al. \cite{aguerri2024fight} & Account Filtering & 05/02/2022-15/03/2022 & -- \\

Toraman et al. \cite{toraman2024mide22} & Event-based keyword search & 24/02/2022-21/03/2022 & Manual annotation\\

Darwish et al. \cite{darwish2023identifying} & Random filter of tweets & not specified & Fact-based annotation \\

Ferdush et al. \cite{ferdush2023detecting} & Keyword Search & 02/2022-05/2023 & Source credibility based annotation \\
\bottomrule 

    \end{tabular}%
    \label{ex:misinfodata}%
\end{table*}%


Bots, or automated accounts, are of particular interest in the dissemination of information, as they distort the perception of public support (or disapproval).
\citet{de2023twitter} used Botometer (previously, BotOrNot) \cite{davis2016botornot} to detect bots in tweets posted by politicians in six major parties in Italy in 2022 after Russia's invasion of Ukraine. 
They discovered that around 12\% of the commenters on these posts were bots, with Giorgia Meloni (who became the Prime Minister of Italy on October 22, 2022) showing the higher percentage of bots (at 15.08\%).
Focusing on pro-Russian tweets, \citet{geissler2023russian} apply Botometer to posters and retweeters of such content, and conclude that ``bots played a disproportionate role in the dissemination of pro-Russian messages'', with ``20.28\% of the spreaders'' being identified as bots, ``most of which were created at the beginning of the invasion''.
Also using Botometer, \citet{zhao2024manufacturing} study a general dataset about the war, and label 23.14\% of the accounts as bots. 
Comparing their activity to that of non-bot (human) accounts using Granger causality test, the authors find evidence of causality in both directions. 
A popular tool for such research, Botometer has unfortunately been unable to update its scores since the closure of the public Twitter API, however historical data is available up to May 2023.\footnote{\url{https://botometer.osome.iu.edu/faq}}
We use this functionality in our study.


Twitter content moderation is an important enforcement mechanism of the platform's terms of service. 
Flagging individual tweets, suppressing their spread, and deleting content have all been possible actions the platform could take to limit the spread of unwanted materials \cite{zannettou2021won}. 
However, suspension of entire accounts has been the most visible and controversial action the platform can take, often referred to as ``deplatforming'' \cite{jhaver2021evaluating}.
Studies of such suspended accounts have shown this approach to be effective in decreasing the overall activity and the toxicity levels of their supporters \cite{jhaver2021evaluating}.
On the other hand, fine-grained tools have been proposed for personalized moderation wherein the user is able to control what is shown in their feed \cite{jhaver2023personalizing}, however these were envisioned for moderating toxic speech, instead of its factual accuracy.
Recently, \citet{pierri2023does} have examined the accounts suspended in a Twitter dataset around the Russo-Ukrainian conflict in the first two months of the conflict. 
They found that Twitter tends to be more proactive in suspending accounts that were created more recently, as well as those that use toxic language and have a higher level of activity. 
However, it is up to the platform whether it chooses to exercise this option.
\citet{pierri2023propaganda} track Russian propaganda and low-credibility content on Facebook and Twitter and find that only about 8-15\% of the posts and tweets sharing links to Russian propaganda or untrustworthy sources were removed.
In this study, we focus on misinformation specifically (as opposed to any toxic or spam content), to measure the extent of moderation Twitter exercised.

\section{Data Collection}

\begin{figure}[!tbp]
    \centering
    \includegraphics[width=1\linewidth]{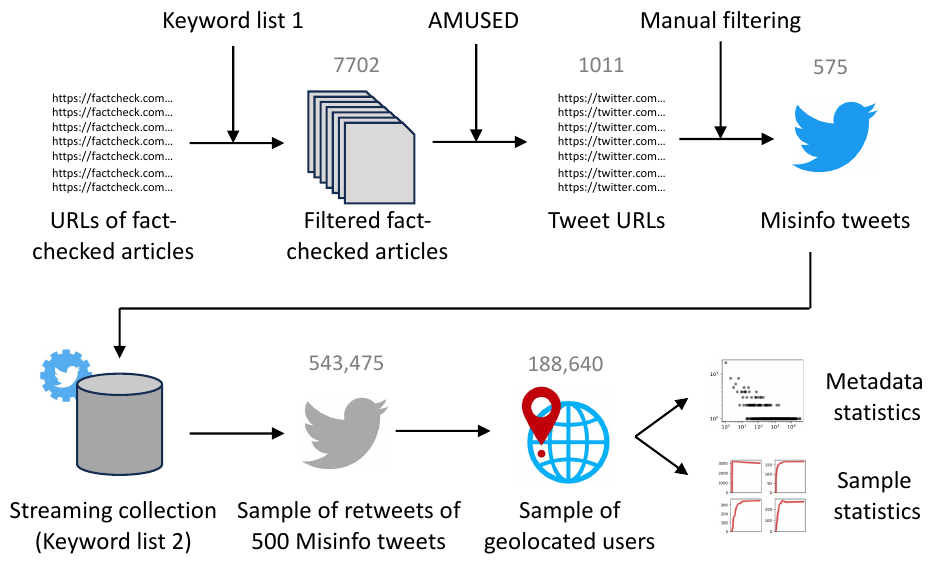}
    \caption{Pipeline diagram for data collection of misinfo tweets and a sample of their retweets.}
    \label{figure:pipeline}
\end{figure}

The data collection took place in two phases (see Figure \ref{figure:pipeline}): the compilation of misinformation tweets mentioned by fact-checking websites and the retrieval of a sample of their retweets. 

\subsection{Fact-checked Misinformation Tweets}

We begin by collating a set of misinformation tweets using the AMUSED framework~\cite{shahi2021amused}, 
which fetches the social media links from the fact-checked articles and assigns a label (verdict of claim in fact-checked articles) to social media posts. 
To do so, first, we collected outlets offering fact-checked articles about the Russian invasion of Ukraine. 
This involved compiling a list of both fact-check data aggregators such Google Fact Check Tool Explorer (GFC),\footnote{\url{https://toolbox.google.com/factcheck/explorer}} Pubmedia,\footnote{\url{https://fact.pubmedia.us/}} European Digital Media Observatory (EDMO),\footnote{\url{https://edmo.eu/}} and fact-checking websites, for instance, Maldita.es.\footnote{\url{https://maldita.es/}} 
For the outlets that offer fact-checked articles on multiple topics (such as Google Fact Check Tool Explorer), we used keyword-based filtering.

We compiled a list of categories as keywords from Wikinews \cite{bruns2006wikinews} for filtering the fact-checked articles; for instance, keywords include ``Ukraine'', ``Russian invasion'' (see Appendix \ref{a:1} for complete list).
Non-English-language articles were translated into English using the Google Translate API.\footnote{https://pypi.org/project/googletrans/}
We then matched keywords in the title or content of the articles to retrieve those relevant to the war, resulting in \num{7702} fact-checked articles from 52 fact-checking websites spanning the period from 24th February 2022 to 31st March, 2023. Some examples of collected fact-checked articles and their transaction are given in Table \ref{tab:examples}.

Next, we selected the tweets that have been identified by various fact-checking organizations as \emph{misinformation}, \emph{fake}, or \emph{misleading} and normalized the label using  \cite{shahi2021exploratory}  
into \emph{false}, \emph{partially false}, \emph{true}, and \emph{other}. 
We fetched hyperlinks referring to Twitter from the fact-checked articles and assigned labels as \emph{false} and \emph{partially false} using AMUSED, which resulted in \num{1010} tweets. 

The last step of the AMUSED framework is manual verification, as fact-checking websites sometimes embed tweets mentioning the debunked claim in the fact-checked articles. 
Hence, we manually cross-checked the tweet linked within fact-checked articles to verify if it was the same tweet debunked in the fact-checking articles and that it wasn't a tweet debunking the original. 
\rev{We obtained \num{1010} misinfo tweets covering 13 months of the Russo-Ukraine conflict. However, around half of them were duplicates (different fact-checking organizations debunked same tweets). 
These tweets were debunked by 41 fact-checking organizations from 25 countries including Ukraine (specifically, by SpotFake.org, which is based in Kyiv) and the majority of International Fact-Checking Network (IFCN) signatories, \footnote{https://ifcncodeofprinciples.poynter.org/signatories} majority of them were debunked by western fact-checkers. 
The maximum number of fact-checking organizations were from the USA, followed by Spain, UK, and India, the same trend was observed in prior studies on claims fetching \cite{shahi2020fakecovid}.  
These tweets are manually debunked by human fact-checkers which is a time-consuming and economically costly process.}
Overall, after removing duplicates we identified \num{575} unique tweets as misinformation with a fact-checked article as background truth.

Once the tweets were identified, an additional data gathering step was performed to verify whether the tweets were visible on the platform.
At the beginning of May, 2024 we collected any possible error message the platform gave for each tweet, including whether the post or the account was deleted, whether the account was suspended, the user limited engagement with the tweet (and it thus was no longer visible), whether the post was unavailable for some other reason, or that it violated the rules of the platform.


\subsection{Misinformation Retweet Sample}
\label{dc:retweets}

To better understand the propagation of the above misinfo tweets, we use a resource created for the study of the Russo-Ukrainian conflict that employed the Twitter Streaming API\footnote{\url{https://developer.twitter.com/en/docs/twitter-api/v1}} to collect a sample of Twitter relevant to the war in the time shortly after the invasion (February 26, 2022) up to the discontinuation of the API (March 14, 2023). 
The keywords used to collect these included the main actors (Putin, Zelensky), locations (Kiev), and ``Ukraine'' translated into at least 50 languages~(see Figure \ref{figure:keywords} for complete list). 
Because the tweets are gathered as they are published \rev{(in a ``streaming'' fashion)}, some that are made private or removed later remain in the set (which is especially important when studying misinformation).
The IDs of the tweets were matched to those in the misinformation set, and both the original tweets and their retweets are extracted. 
This filtering resulted in \num{543475} tweets (posted by \rev{\num{381186}} users), with \rev{500} unique misinfo tweets (posted by \rev{428} unique users) having at least one match. 
Judging by the metadata of these 500 tweets, they received \num{877579} retweets by the end of data crawling (which might change in the present if a tweet exists), thus our sample retweets contain around 62\% of these retweets.
In summary, we had two sources of information: (1) metadata in the captured data objects (such as number of retweets a tweet has received, or information about the user), and (2) information about a sample of the retweets (including when retweets occurred, and users who have retweeted it).

\begin{figure}[t] %
    \centering
    \includegraphics[width=\linewidth]{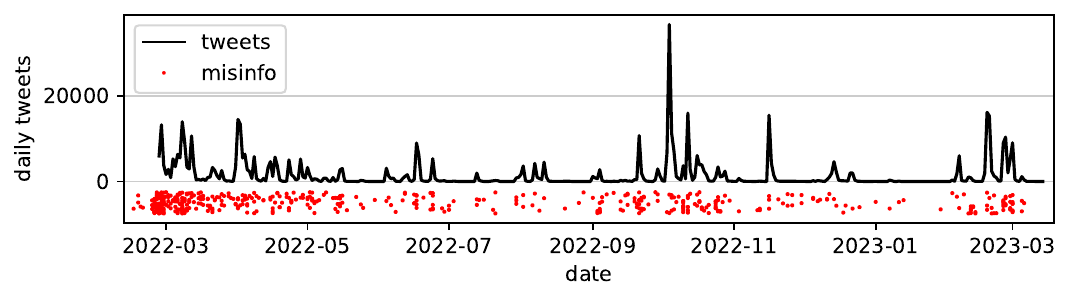}
    \caption{Daily number of misinfo tweets (in red, jittered vertically) and their retweets in our sample (black line). Four instances predating plot time frame excluded.}
    \label{fig:volume}
\end{figure}

The volume of the data and the instances of misinfo tweet postings are shown in Figure \ref{fig:volume}. 
Not shown are four instances of misinfo tweets posted on January 18 and 24, 2022, and January 17, and September 25, 2019. 
We find a large portion of the misinfo tweets (\rev{53\%}) was posted within the 3 months following the invasion, with additional concentrations around October, 2022 and March 2023. 
Although we do not possess the data before February 26, previous works on the subject suggest the conversation around the conflict was minimal before the invasion \cite{pierri2023propaganda}.

\subsection{User Geolocation}
\label{dc:location}

We supplement the information about the users who have posted misinformation (misinfo users) and those who have retweeted them by examining the Location field in their profile. 
We map this field to a large database of geographic locations GeoNames\footnote{\url{https://www.geonames.org/}} \cite{mejova2021youtubing}.
As the field does not have any format constraints, the users may enter anything into it, thus we manually check the top 3000 most popular location matches to make sure matches which are obviously not locations (such as ``on my bed'') are not mistakenly matched to a real location. 
We were able to geolocate \rev{294} (\rev{69\%}) of the users who posted misinfo tweets and \rev{\num{188640}} (\rev{50\%}) of those who retweeted them \rev{(this proportion is higher than geolocation success rate of a more general subset from \cite{mejova2021youtubing}, which was at 42.4\%)}. 
We emphasize that the location field is not verified in any way, thus it may be manipulated by users to appear to be from a certain location to intentionally mislead the users of the platform.

\begin{figure*}[!htbp]
    \centering
    \subfloat[\rev{Users sending misinfo}]{\includegraphics[width=0.47\linewidth]{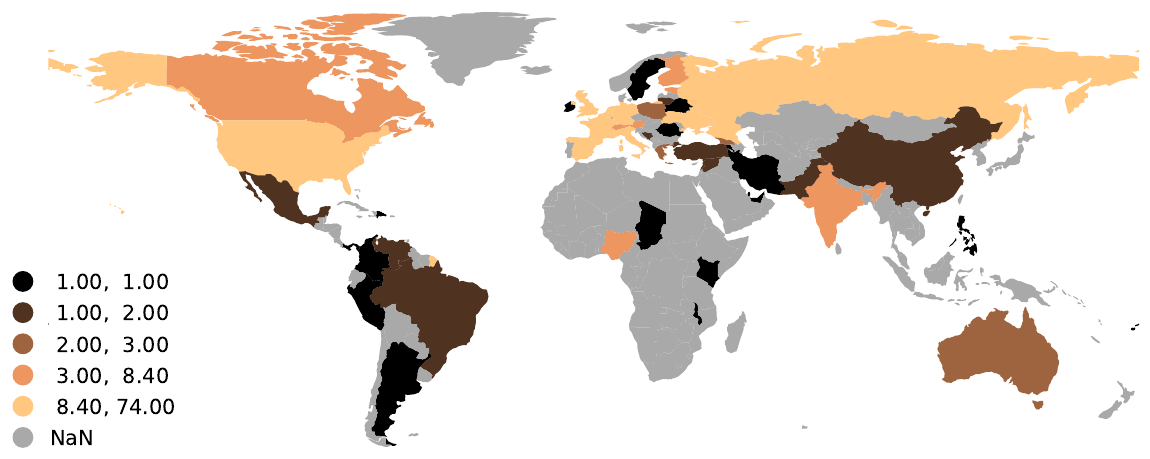}}\hspace{0.5cm}
    \subfloat[\rev{Users retweeting misinfo}]{\includegraphics[width=0.47\linewidth]{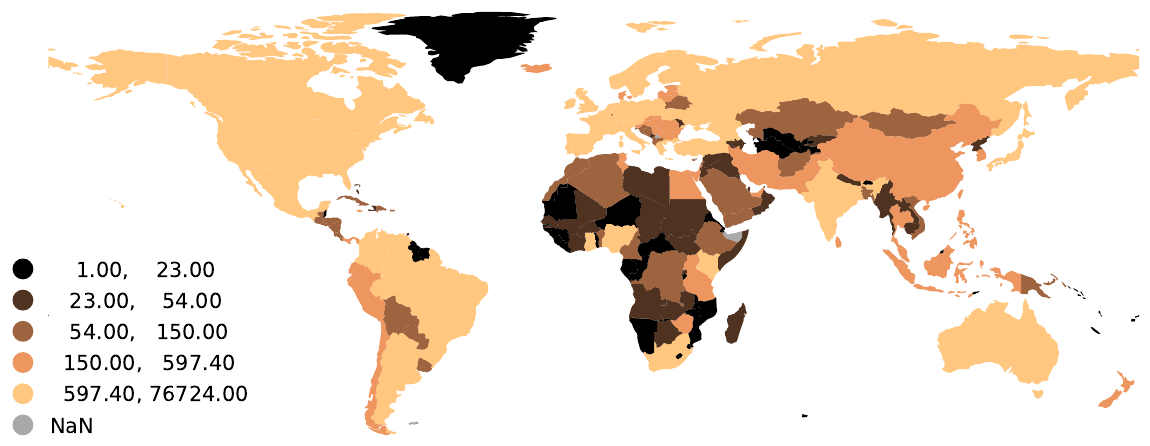}}
    \caption{Geographical distribution (map created with GeoPandas; boundaries are not exact and NaN is zero) of users (a) who have posted misinformation tweets and (b) those who have retweeted them. Colors are segmented by quintiles.}
    \label{fig:maps}
\end{figure*}

\rev{To summarize}, out of 575 tweets identified as misinformation, our sample contains \rev{500}, which were posted by \rev{428} users, \rev{294} of whom were geolocated. 
Our retweet sample captured \rev{\num{543475}} retweets posted by \rev{\num{381399}} users, \rev{\num{188640}} of whom were geolocated. 
Below, we explore the extent to which these \rev{500} instances of misinformation spread across the world, and what actions the platform has taken concerning them. 
We provide the misinfo tweets, their metadata, and the IDs of their retweets to the research community.$^4$ 

\section{Results}


\subsection{Sources of Misinformation}

First, considering the users who have posted misinfo tweets, we find that a surprising number of them, \rev{39.4\%}, are verified accounts (and \rev{60.6\%} are not verified).
Note that the dataset spans the time before Twitter started paid account verification, such that only selected accounts were awarded these tags after verification from the Twitter team \cite{shahi2021exploratory}. 
Around \rev{11.7\%} of accounts were created after the start of the invasion, however misinfo tweets posted by these new accounts received much fewer retweets (on average \rev{66.7\%} fewer) compared to accounts created before, pointing to the importance of the accumulated social capital. 
When we consider the countries users posting misinfo tweets have declared themselves to be located in (see Figure \ref{fig:maps}(a)), most are most likely to be geolocated to \rev{US (25\%), followed by Ukraine (12\%), Great Britain (7\%), France (5\%) and Russia (5\%)}.
Finally, \rev{we used Botometer (version X) \cite{yang2020scalable} to calculate the bot score of each Twitter account}; out of \rev{428} users, \rev{378} were still available. 
We manually checked the \rev{Twitter accounts through Botometer user interface.\footnote{https://botometer.osome.iu.edu/}} and analyzed 30 accounts that scored more than 2 (out of 5) on the Botometer scale. Out of these, we identified 12 as potential bots or having suspicious activities, and all of them were not verified.

Interestingly, the account responsible for most of the tweets in our misinfo tweet set (7 misinfo tweets in total) is @NEXTA, self-described as ``The largest Eastern European media. To let the world know.''
Its Patreon page (linked to in the Twitter description field) says it is a media channel from Belarus, having a stance in opposition to the President of Belarus, Aleksandr Lukashenko. 
The Twitter account was unavailable at the time of writing.
The second account responsible for most tweets (in this case, 5 misinfo tweets) is the self-described ``Ministry of Foreign Affairs of Russia (Official account)'', which was a verified account at the time of the collection (also now available).
Tying for third place (contributing 4 misinfo tweets) are Dmitry Polyanskiy, ``First Deputy Permanent Representative of Russia to the UN'' (account available), and Maxar Technologies, a space technology company (also available). 
In the list of accounts contributing 3 misinfo tweets we find a Ukrainian newspaper The Kyiv Independent, a columnist at The Daily Beast Julia Davis, and the official channel of the UK Ministry of Defense.
Thus, we find a wide variety of sources, including journalists, political actors, and even data providers (a satellite company), illustrating vividly that false or misleading information can come from a variety of sources.

\subsection{Content of Misinformation}

Turning to the \rev{500} misinfo tweets posted by these users, as the list of fact-check articles spanned many different countries, our misinfo tweet set spans 17 languages.
\rev{Among these, most are in the English language (71\%), followed by Spanish (7.8\%), Italian (5.2\%), and French (4.2\%). Only 5 tweets were in Ukrainian (1\%) and 1 in Russian (0.2\%).}
Interestingly, most misinfo tweets identified as originating in Russia used the English language in their text (\rev{15 out of 20}), similarly to Ukraine: \rev{36 out of 44} tweets are in English.
 


\rev{In order to structure our analysis, we use the narrative framework~\cite{de2021internet}, which builds the scaffold of the narrative by employing the main actors of the story and their role within it. 
Given the two sides of the conflict, we identify Russia and Ukraine as two main actors (both sides include the mentions of the nations, as well as their politicians and peoples).
We then combine the two role dimensions of the narrative framework -- moral quality (benevolent vs.~malevolent) and power (strong vs.~weak) -- into valence stances that are promoted in the post. 
For instance, posts aiming to portray Ukraine as weak or malevolent are labeled as anti-Ukraine, and those aiming to portray it as strong or benevolent as pro-Ukraine (similarly, for Russia).
The annotation was carried out by one researcher and cross-checked by another and we finalized the annotation once we reached an agreement.
Cohen's Kappa was computed on a random set of 20 tweets, resulting in $\kappa=0.76$, indicating a good inter-coder agreement.
Out of the 500 tweets, 188 (37.6\%) were labeled as anti-Ukraine, 94 (18.8\%) as pro-Ukraine, 90 (18\%) as anti-Russia, and only 18 (3.6\%) as pro-Russia.
The remaining 110 tweets (22\%) were labeled as Other, as the focus of their narratives were not the two sides above (in these, often NATO, US, and the media were mentioned).}

Examples of tweets with each position can be seen in Appendix Figure \ref{fig:examples}. 
An example (Figure \ref{fig:examples}(a)) of the \textit{pro-Ukraine narrative} is a tweet in which a photo of an attractive woman dressed as a soldier is falsely claimed to be the wife of Ukrainian President Zelensky. 
An example (Figure \ref{fig:examples}(b)) of a \textit{pro-Russia narrative} tweet in a reference to a purported expression of support by the Uganda’s President Yoweri Museveni of the Russian President Putin's actions in Ukraine.  
The \textit{anti-Ukraine narrative} includes tweets suggesting Ukraine was developing biological military weapons in secret bio-labs (Figure \ref{fig:examples}(c)). 
Finally, \textit{anti-Russia narrative} often emphasizes the cruelty of Russia during the conflict, such as (debunked) reports of using military vehicles to destroy civilian cars (Figure \ref{fig:examples}(d))). 
Often, the content of narrative is repetitive, and multiple variants of these misinfo tweets were posted by different users.

Especially common are various personal attacks of significant figures. 
For instance, Ukrainian president Volodymyr Zelenskyy has been said to use a body double during the public appearance (implying weakness), or to make money from western countries in the name of war (implying malice).
On the other hand, the Russian president Vladimir Putin has been accused of using a green screen for public appearances (implying weakness).
Additional themes concerned the nations themselves, such as showing people of Ukraine going to the beach and marrying during the war, diminishing the impacts of the war on the civil population; whereas other posts focus on the military action, claiming damage and describing weaponry on both sides; yet others consider the role of the world political leaders in escalating or deescalating the war, e.g.~then U.S. President Joe Biden or the Indian Prime Minister Narendra Modi.
Finally, some posts address various consequences of the conflict, including possible discrimination against refugees on border crossings, economic sanctions against Russia, and the (often mis-attributed) commentary on the conflict by notable individuals.

Finally, to our surprise, two of the oldest tweets were originally published in 2019 and were ``repurposed'' during the Russo-Ukraine conflict. 
The first such tweet was posted by a Russian journalist about the revocation of sanctions in 2019 after the 2014 Russia-Ukraine war. 
It was debunked by \textit{Maldita.es} stating, ``Although it is Ukraine, the image dates back to at least 2019 and is not linked to the current Russian bombing''. 
The second tweet implied that the Ukrainian people were gathering during bombings to pray, however the image was from 2019 and comes from the website of the International Mission Board, a Baptist missionary organization.


\vspace{-0.1cm}
\subsection{Reach of Misinformation}

\begin{figure}[!tbp]
    \centering
    \vspace{-0.3cm}
    \includegraphics[width=.63\linewidth]{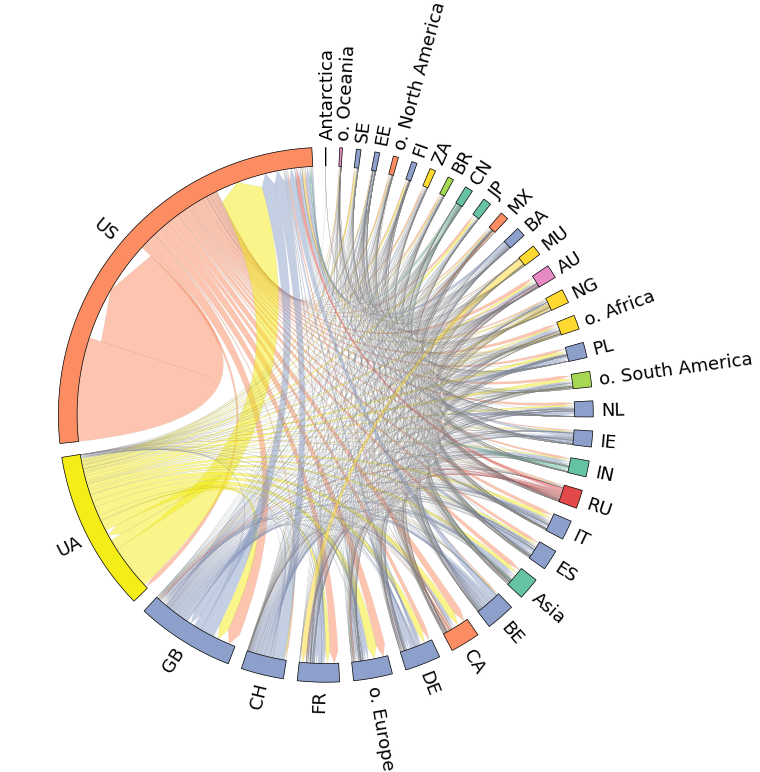}
    \caption{Tweet flows between countries (smaller countries aggregated by continent, with ``o.'' meaning ``other'').\vspace{-0.2cm}}
    \label{fig:circle}
\end{figure}

Considering the reposters of these posts, at least those who were captured in our dataset (recall that it is a sample), we found \rev{\num{381186}} unique users who have reposted this information, \rev{\num{188640}} of which we were able to geolocate.
The geographic distribution of these retweeters of the misinfo tweets can be seen in Figure \ref{fig:maps}(b).
Both original posters and retweeters are more likely to declare themselves to be in the US, Western Europe, and Russia, but retweets also reach much of South America and Eastern Europe. 
Specifically, \rev{19.5\%} of the retweeters were in the US, \rev{3.4\%} in Great Britain, 2.2\% in France, 1.8\% in Canada. Only 1.5\% were in Ukraine and 0.3\% in Russia.
Figure \ref{fig:circle} shows the flow of misinfo tweets between countries, with smaller ones aggregated by continent (``o.'' meaning ``other''), colored by continent (except Ukraine and Russia).
We find that, after the U.S., accounts claiming to be from Ukraine make up some of the largest volume \rev{of misinfo tweets, many of which are then retweeted by accounts in the U.S.}
\rev{Interestingly, these Ukrainian users retweet very little content from other countries.}

\begin{figure}[!htbp]
    \centering
    \subfloat[Number of retweets]{\includegraphics[width=0.49\linewidth]{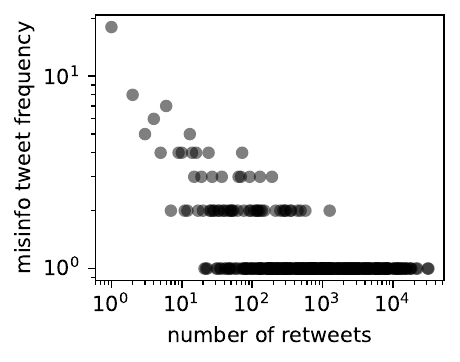}}
    \subfloat[Number of days]{\includegraphics[width=0.49\linewidth]{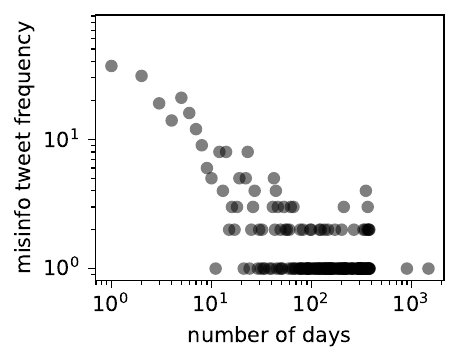}}
    \caption{Number of retweets per misinfo tweet and the number of days until the last retweet.}
    \label{fig:rtstats}
\end{figure}

However, the above percentages hide the substantial number of accounts who have retweeted these misinfo tweets: according to the tweet metadata, the 500 misinfo tweets under our consideration were retweeted a total of \rev{\num{877579}} times, likely reaching views into many millions. 
Figure \ref{fig:rtstats}(a) shows the frequency statistics of the number of retweets.
Only \rev{18} tweets were never retweeted, the rest display a long tail, with the maximum reaching to \rev{\num{31941}}. 
Due to this heavy tail, the average number of retweets is \rev{1755}, however the median is only \rev{283}. 
When we consider the duration of the retweet cascade (the number of days between the original post and the last retweet), we find a similarly long-tailed distribution in Figure \ref{fig:rtstats}(b).
In this case, we find the two posts from 2019 having resurgent retweet activity which result in the last retweet happening from 897 to 1473 days after the original post. 
Upon excluding these two tweets, we find that the average cascade duration is \rev{74.5} days, however the median is only \rev{12} days, and 25th percentile only 2. 
\rev{These numbers hide, however, just how explosive the sharing is immediately after the posting of a tweet: 90\% of the tweet's retweets on average come within the first 51 hours (median of 28 hours), pointing to the necessity of timely action on the part of moderators. }

Among the retweeting users captured in our dataset, the vast majority (\rev{77.6\%}) retweeted only one of the misinfo tweets, suggesting that the material reached a broad audience.
On the other hand, \rev{73} users have retweeted 20 or more of the misinfo tweets in our set, with the top retweeting user retweeting 71 misinfo tweets (that we could capture).
This user has no informative bio or link, geolocates to Australia, and has created the account in 2013. 
The Botometer score for this user is low (1.0), indicating it is unlikely to be a bot. 
On average, these 73 users have a Botometer score of 1.6, with a maximum score of 4.1, and only 4 accounts were created after the invasion.
We manually annotated 24 accounts with a Botometer score of 2.0 or more and found that 13 accounts looked as bots or suspicious (these accounts were either newly created or, and the number of followers is less and in the short span of time, account posts or retweet a thousand posts sometimes only on one topic such as Russo-Ukraine conflict), and 5 had high Botometer scores, but the accounts were deleted.
Note that the results of this small annotation effort suggest that Botometer scores may need to be manually verified when applying to new data. 
Thus, although some may be automated, most are likely to be operated by human users who were on the platform before the conflict escalated.

\begin{figure}[t]
    \centering
    \subfloat[Number of retweets]{\includegraphics[width=0.49\linewidth]{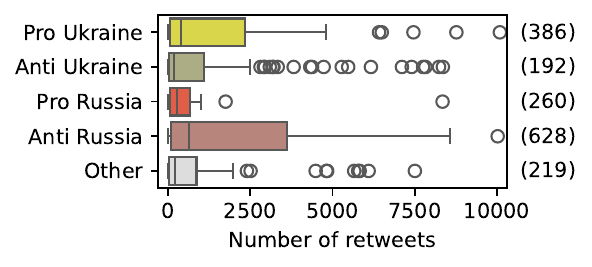}} 
    \subfloat[Number of likes]{\includegraphics[width=0.49\linewidth]{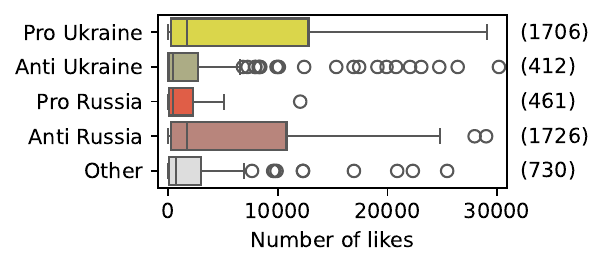}}
    \caption{Distributions of the number of (a) retweets and (b) likes for misinfo tweets with various leanings, with class medians in parentheses (outliers cropped for clarity).}
    \label{fig:proanti_engagement}
\end{figure}

\rev{When we consider the engagement with these tweets for each narrative leaning separately, we find a definite bias towards two stances: pro-Ukraine and anti-Russia (see Figures \ref{fig:proanti_engagement} (a,b)).
Anti-Russia misinfo tweets receive the most retweets (a median of 628), although in terms of likes anti-Russia stance rivals pro-Ukraine one (at around 1700 likes at the median).
Instead, the stances anti-Ukraine and pro-Russia receive much less engagement, at around 200 retweets and 450 likes, about the same retweet amount as the tweets annotated as having "other" stances (though these receive more likes).
This heightened attention to pro-Ukraine, anti-Russia view points to the potential support of this viewpoint by the platform's (largely Western) users.}


Finally, we consider characteristics of the user to model the popularity of the content they posted using regression analysis.
Note that here we are not attempting to develop a robust prediction algorithm, but use the regression analysis as an explanatory device. 
Table \ref{tab:regression} shows the coefficients of two linear regression models, one predicting the number of retweets and another likes that a misinfo tweet received, both log-transformed.
The independent variables include the log-transformed number of followers the posting user has at the time of the posting, whether the account is verified, whether it was created before the invasion, whether the user is geolocated to one of the most popular detected locations (US, UA, GB, RU), whether the language of the tweet is English, and finally whether eventually the account was found to be unavailable (either suspended or deleted by a user).
The above list of features is the result of feature selection using the Variance Inflation Factor (VIF), selecting features until VIF score for each is less than 6.
This process removed most of the activity features, including number of lifetime tweets posted by a user, age of the account in terms of days ~as they introduce multicollinearity. 
We find that the most predictive variable is the number of followers, which makes intuitive sense, since a larger audience makes the spread of information more likely.
Interestingly, whether the account is verified does not significantly relate to popularity. 
Further, whether the account was created before the invasion also does not have a significant coefficient: the effect of this variable is likely taken up by the number of followers, as older accounts are more likely to have accumulated more social capital.
Out of the geolocated countries, we find that the tweets by users claiming to be in Ukraine have a higher chance to be retweeted and liked. 
Whether the tweet is in English is not associated with more attention.
Finally, whether eventually the account was suspended or deleted has no linear relationship with engagement (putting in question the effectiveness of these measures, as we see in the next section). 
Note that we also ran a similar model with 463 tweets for which we were able to obtain a Botometer score (omitted for brevity), and the score had no significant relationship to engagement.


\begin{table}[!tbp]
    \centering
    \caption{Linear regression models predicting number of retweets and number of likes (both log-transformed).}
    \label{tab:regression}
\begin{tabular}{lrlrl}
\toprule
Feature & \multicolumn{2}{c}{Retweets} & \multicolumn{2}{c}{Likes}   \\
\midrule
constant & 0.454 &  & 0.944 & . \\
followers$_{log}$ & 0.445 & *** & 0.517 & *** \\
verified & -0.406 &  & -0.345 &  \\
country: US & 0.230 &  & 0.203 &  \\
country: UA & 0.943 & * & 1.127 & * \\
country: GB & -0.141 &  & -0.121 &  \\
country: RU & -0.184 &  & -0.445 &  \\
language: EN & 0.323 &  & 0.443 &  \\
unavailable & 0.347 &  & 0.386 &  \\
\midrule
& \multicolumn{2}{l}{$adj R^2 = 0.295$} & \multicolumn{2}{l}{$adj R^2 = 0.327$} \\
\bottomrule
\multicolumn{5}{l}{\footnotesize *** $p<0.0001$, ** $p<0.001$, * $p<0.01$, . $p<0.05$} 
\\
\end{tabular}
\end{table}

\subsection{Platform Response}
\label{sec:platform_response}

\begin{table}[!htbp]
    \centering
    \caption{Status of tweets and accounts posting them.}
    \label{tab:moderation}
\begin{tabular}{llrr}
\toprule
Tweet status & Account status & Freq. & \% \\
\midrule 
Available & Available & 408 & 81.9 \\
Deleted post & Available & 30 & 6.0 \\
Suspended account & Suspended & 26 & 5.2 \\
Deleted post & Doesn't exist & 12 & 2.4 \\
Available & Doesn't exist & 10 & 2.0 \\
Suspended account & Doesn't exist & 7 & 1.4 \\
Deleted post & Suspended & 1 & 0.2 \\
Limited engagement & Available & 1 & 0.2 \\
Limited engagement & Protected view & 1 & 0.2 \\
Unavailable post & Withheld & 1 & 0.2 \\
Violated rules & Available & 1 & 0.2 \\



\bottomrule
\end{tabular}
\end{table}

\begin{table*}[!htbp]
\centering
    \caption{Status (percentage) of tweets according to their leaning, normalized within each category.}
    \label{tab:moderation_leaning}
    \begin{tabular}{lrrrrrr}
\toprule
Leaning & Available & Deleted & Limited engagement & Suspended account & Unavailable & Violated rules \\
\midrule
Pro-Ukraine & 85.1 & 10.6 & 1.1 & 3.2 & 0.0 & 0.0 \\
Anti-Ukraine & 78.7 & 9.6 & 0.5 & 10.1 & 0.5 & 0.5 \\
Pro-Russia & 77.8 & 5.6 & 0.0 & 16.7 & 0.0 & 0.0 \\
Anti-Russia & 90.0 & 8.9 & 0.0 & 1.1 & 0.0 & 0.0 \\
Other & 88.2 & 5.5 & 0.0 & 6.4 & 0.0 & 0.0 \\
\bottomrule
\end{tabular}
\end{table*}

\begin{figure}[!htbp]
    \centering
    \includegraphics[width=.99\linewidth]{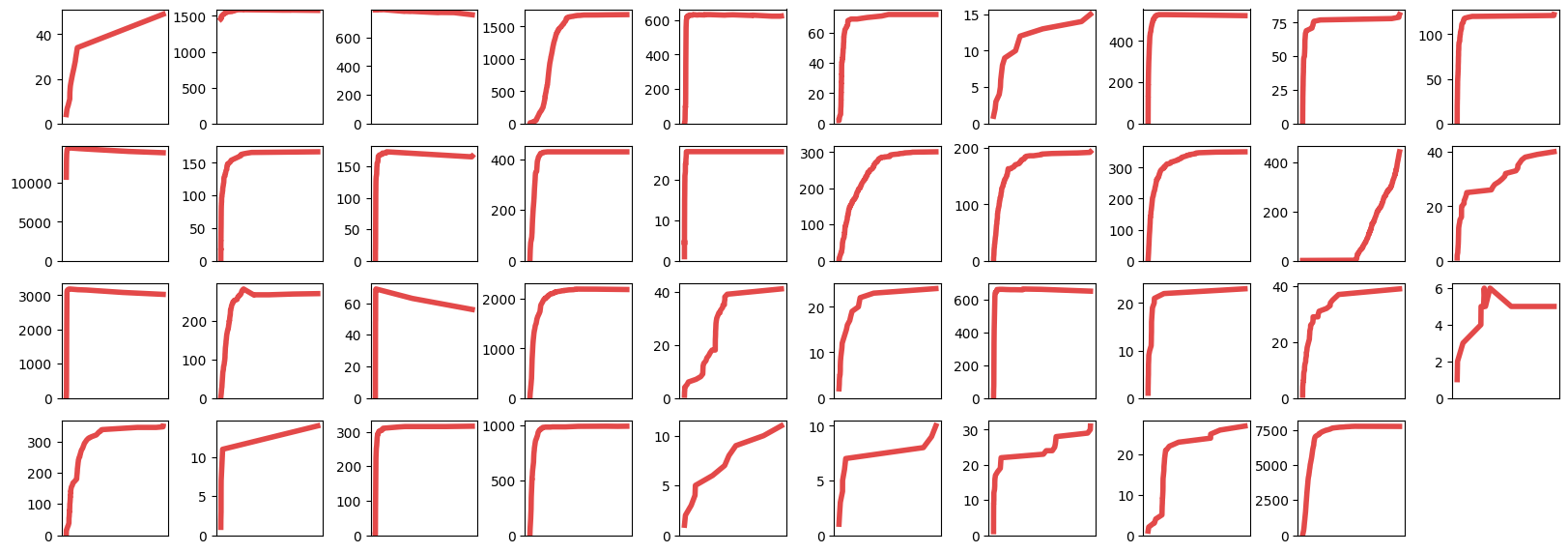}
    \caption{Number of retweets over time ($x$ axis is hidden for clarity) for misinfo tweets that were moderated (for which over 5 tweets were captured).}
    \label{fig:moderated_cascades}
\end{figure}

Over a year after the last of the posts in our dataset was posted, the misinfo tweets and the posting users were re-queried on the Twitter interface, and as Table \ref{tab:moderation} shows, \rev{84\%} of the posts remain available on the platform. 
We have observed two measures from Twitter as a platform. 
First, the platform \textit{suspended \rev{27} accounts}, however we could extract the location of 10 users who posted these tweets. 
It suspended 7 accounts from Europe, 2 from North America, and one from Asia. 
Another measure platform could take is \textit{flagging tweets as violating rules}, but only one tweet was flagged as violating rules, not the account. 

From the users' side, \rev{29} accounts were deactivated or renamed by the users themselves; all were unverified accounts. 
Note that in the case in which a post is still available, and the account is not, the platform redirects visitors to a tweet on an account with a different handle (and different numerical ID), thus obscuring the fact that the original version of the account has been removed.
Two tweets were restricted in their engagement, one (unverified) user protected the views of all of their tweets, and one verified user withheld the account from the researchers' country.
Overall, the measures taken by the platform affected about a sixth of the misinfo tweets in our set. 
This finding echoes that of a recent study of Russian propaganda and low-credibility content on Facebook and Twitter \cite{pierri2023propaganda}, who find 8-15\% of such posts and tweets being removed.

Although the platform does not specify when the moderation action was undertaken, we delve into our dataset for evidence of its impact.
Figure \ref{fig:moderated_cascades} shows the number of retweets we witness over time (which is the $x$ axis, which is hidden for clarity of presentation).
We display only cascades for which we have more than 5 retweets captured in our dataset.
Examining the trend lines, we find that, by the last retweet we witness, the number of retweets has stabilized, and in some cases it is decreasing (perhaps after the claim is debunked).
\rev{Indeed, most of the retweeting activity happens within the first 2 days of the tweet's posting.}
\rev{We also check the status of the tweets with respect to the narrative leaning in Table \ref{tab:moderation_leaning}.
Interestingly, the anti-Russia posts were most likely to be still available at the time of the analysis (at 90\%), compared to, say, pro-Russia or anti-Ukraine  (both at 80\%, comparison with anti-Ukraine is statistically significant at $p<0.05$ using $t$-test).
This tendency may reveal slight disregard by the Western-owned company to take action when misinformation is unfavorable to a perceived adversary (Russia).}

\rev{Further, we computed the time taken by fact-checkers in debunking misinfo tweets as the difference between the tweet posted and the date of fact-checked articles published. 
For difference calculation, we ignored the old repurposed tweets such as the tweet created in 2017. The median time taken by fact-checkers to debunk misinfo tweets is 6 days, however a long tail of delayed fact-checks results in a much larger average of 30 days. 
As it takes a median misinfo tweet in our collection 28 hours (around 1.6 days) to reach 90\% of its all-time retweet popularity (on average, 51 hours or 2.1 days), meaning that the vast majority of tweets have already reached most of their audience by the time a debunking article is published.}

In summary, we find a systemic mismatch between the propagation dynamics of the content, the fact-checking, and the actions of the platform that prove the current misinformation efforts ineffective.

\section{Discussion}


There are many complexities in the nature of ``misinformation'' that make its study a challenging task.
The definition of ``fake news'' or false information lies on a spectrum: fact-checkers often flag tweets that are only partially true, misleading, or taken out of context. 
For instance, the tweets posted in 2019 may have been appropriate at the time, but were ``repurposed'' during the wartime in 2022, in the content that made incorrect implications.
Further, a user may attempt to correct the original post with additional information, but it is excluded in the subsequent resharing of the original. 
However it may happen, and for whatever possible existence or absence of malicious intent, our data illustrate that such content does not necessarily originate with bots or newly created accounts. 
Instead, it may come from the official accounts of ministries, politicians, and news agencies, all of which have motivations for participating in the shaping of the narrative and being first to break a story, with less incentive to first do due diligence, or carefully consider the way the information may be misperceived. 
Although we do not dismiss the presence of intentional disinformation campaigns, we found a variety of narratives in the tweets the fact-checkers cited.

It may be especially important to track the narratives of tweets during the Russo-Ukrainian conflict, including those possibly misinforming their audience, due to the crucial role popular opinion may play in the relevant policy decisions. 
Ukraine requires the support of its European allies to maintain its military and civilian infrastructure \cite{trebesch2023ukraine}, and degrading this support would be beneficial to Russia.
The information does not even need to be false to sway public opinion, and many other discursive approaches can be used to portray a certain narrative by showing a biased selection of news or framing them in particular light, as agenda setting literature has amptly described \cite{scheufele2007framing}.
Thus, narrative-based tracking may be as important as that based on factual accuracy when tracking information campaigns designed to mislead.


Still, despite the large numbers of retweets and potential audience in the millions, it is not clear whether this content substantially swayed public opinion. 
For instance, \citet{hameleers2023mistakenly} conducted a 19-country survey on the perception of news around the Russo-Ukrainian conflict.
They find that people around the world are already skeptical of the information they receive about the conflict, and ``are more likely to attribute false information to deliberative deception than to a lack of access to the war area or inaccurate expert knowledge.''
The process of translating public opinion to governmental action is further complicated by electoral and EU politics \cite{krastev2024wars}. 
Turning toward the platform, and its users, below we outline several design implications of our findings that may improve the quality of information available to its users, during this conflict and in general.

\subsection*{Design Implications}

\textbf{Global Nature of Misinformation} Unsurprisingly, tweets from users with high follower or friend counts receive more retweets.
Still, despite the methodological ease of focusing on the ``superspreaders'' of misinformation, the fact that 77.6\% of the retweeters were responsible for only one misinfo retweet suggests there is a broad interest around such content, which is supported by existing literature on the importance of peripheral participants in the spread of information \cite{barbera2015critical}.
These findings contrast those of an earlier analysis of tweets about the war in the first 150 since the invasion by \citet{robinson2023casual}, who find that only 49.5\% of their dataset was produced by users who have engaged only once. 
It is possible that the discrepancy stands in the linguistic scope of the datasets: whereas only 71\% of our dataset is in English, Robinson queries for only two words, both in English.
Thus, we urge the fact-checking community to diversify their efforts away from English, and set lower thresholds for popularity.


\noindent \textbf{Automated Activity} Despite popular fears, in our data we do not find a large population of bots, finding 12 out of 428 users posting misinformation to be possible bots and 13 out of 73 of those who retweeted the most of it being tagged as suspicious (though recall that only those which had a Botometer score of larger than 2 were examined). 
These rates are much lower than those obtained in recent previous work \citet{pierri2023propaganda}.
However note that, unlike in much previous work, we manually labeled the accounts which were flagged by Botometer.
The fact that we did not manually identify all of them as bots suggests that the system may need to be adjusted to new applications and scenarios. 
Further, if indeed real people are behind these accounts, it would be best if the platform communicated with them during the process of moderation. 
Previous research has found that moderation actions could result in negative emotional responses \cite{ma2023users}, perceptions of unfairness if the decision process is not clearly explained \cite{nurik2019men}, 
and attempts to circumvent future triggering of flagging of their content \cite{gibson2019free}. 
Moderation steps may not need to be the removal of the content, but tagging and adding links to reputable information sources \cite{sharevski2022mis} could provide a context for the future development of the narrative.




\noindent \textbf{Persistent User-Content Linking} Perhaps the most damning finding of this study is the lack of effective response by the platform to moderate tweets: 84\% of them were still available (and all but one unflagged) to view more than a year after they were posted. 
Although we do not presume to advocate in favor of the harshest sanctions against all posters in our dataset, as some may be misunderstandings or improper contextualization. 
Still, our analysis revealed that it is possible for the tweet to be available, while the original posting user account would show ``Does not exist'' error.
We hypothesize that in such cases, the user is able to change the identity of their account without losing the link to their historical tweets (we find 11 such cases in our data). 
This system behavior may obscure the identities of the original posters of misinformation, and stifle the tracking of accounts of potential offenders.
We urge the platforms to preserve historical URLs of content, such that it may be analyzed and verified, even if the user has changed their account metadata.

\noindent \textbf{Timeliness of Moderation} This study shows that the temporal aspect of the spread of posted information, and the necessary triage, research, and publishing involved in the fact checking process, preclude the actual catching of new misinformation that may appear on the platform. 
Numerous automated systems have been proposed to identify possible misinformation \cite{zhou2020survey}. 
Such systems may consider the veracity of the information it presents, the writing style, its propagation patterns in the social network, and the credibility of its source.
The latter is often used as a ``distant supervision'' method wherein a list (or a ``gazetteer'') of domains known to sometimes or often publish low-credibility content is used to label all content from that domain as having ``low-credibility'' \cite{lin2023high}.
However, as we find in our data, many official and otherwise reputable sources may share questionable information, making such a broad approach either too restrictive (if a high number of poor-quality posts are necessary for the inclusion in such a gazetteer), or too broad (if only one post mars the reputation of a source). 
In any case, even if such systems performed with a high accuracy, without expert (human) oversight, the application of automated ``fake news'' detection systems may result in an overzealous (or even biased) limitation of free speech, bordering on censorship. 

Barring a substantial increase in the resources available to fact-checking organizations or automatically tagging potentially offending content, it is unlikely that we will witness a system that will be able to intercept a spreading rumor within the first couple of days (as seems to be necessary for Twitter).
It is then even more important that ameliorative actions are taken when the fact-checking process results in a report. 
Even if the content has already reached a large audience, it should be labeled as fact-checked, with the proper citation to the article. 
This may result not only in the contextualization of the information for those who will encounter it in the future, but also as a form of feedback to the poster. 
If possible, a way for the users to find out more about the action should be provided, if they have a question about the moderation action \cite{ma2023users}. 

\subsection*{Limitations}

This study has several notable limitations.
Although the effort to collect a diverse sample of fact-checked articles has been extensive, it is unlikely that we covered all resources for the Western world. 
The fact that the set of misinfo tweets we study here are mostly in English suggests that other European languages may still need further effort.
Even if the fact-checking resources were exhaustive, only fairly popular pieces of information are likely to be checked, biasing our selection to those pieces of misinformation that received much attention, possibly leaving out less popular posts which never made it on the fact-check ``radar'' (recall that \rev{39.4\%} of the misinfo accounts were verified, despite being a minority of all accounts on the platform).
The Western bias may have also contributed to the reason why we do not see much attention to the posts from users identifying themselves to be in Russia (although their absence can also be explained by the blocking of Twitter in Russia,\footnote{\url{https://www.theguardian.com/world/2022/mar/04/russia-completely-blocks-access-to-facebook-and-twitter}} and accounts simply not declaring themselves from there). 
An additional methodological limitation is the use of the Location field in the user's profile for geolocation. 
As the field is manually filled by the users, there is no verification to this information (thus we try to be careful to report it as ``location user claims to be in'' instead of ``location user is in''). 
Still, the approach has been used in existing literature \cite{paoletti2024political}, and has been shown to be largely accurate when compared to the small amount of geolocation information available.
We have considered normalizing the results using the national Twitter platform penetration estimates, however since they are not available for all countries, we chose to report the raw numbers, lest we introduce additional bias.
In the case of this conflict, an additional difficulty presents itself in nation-level limitations on information access and self-expression; for instance in Russia, Twitter was blocked within weeks of the invasion.
Further, there is the aforementioned limitation on the data accessibility due to the public Twitter API access being discontinued in 2023.
Even with an academic subscription, it is now impossible to collect datasets on the scale previously done (such as the original streaming data used in this study). 
Thus, the authors of this work share the misinfo tweet data, and the IDs of the retweets, as allowed by the platform Terms of Service. 
For collaboration concerning the entire dataset, we ask the reader to contact the authors directly. 
As Twitter users are known to be younger, more educated, and wealthier than general public in the U.S.,\footnote{\url{https://www.pewresearch.org/internet/2019/04/24/sizing-up-twitter-users/}} the way misinformation is posted and propagated there may not reflect the situation on other platforms such as Truth Social \cite{gerard2023truth}.

\noindent \textbf{Positionality} \\
Finally, the authors acknowledge their own positional bias in the analysis of the content, coming from a Western-centric perspective. We encourage researchers from diverse backgrounds to contribute to this and wider study of misinformation.

\bibliographystyle{ACM-Reference-Format}
\bibliography{ukraine_misinformation}


\appendix

\section*{Appendix}

\begin{figure}[!h]
    \centering
    \subfloat[Pro-Ukraine]{\includegraphics[width=0.24\textwidth]{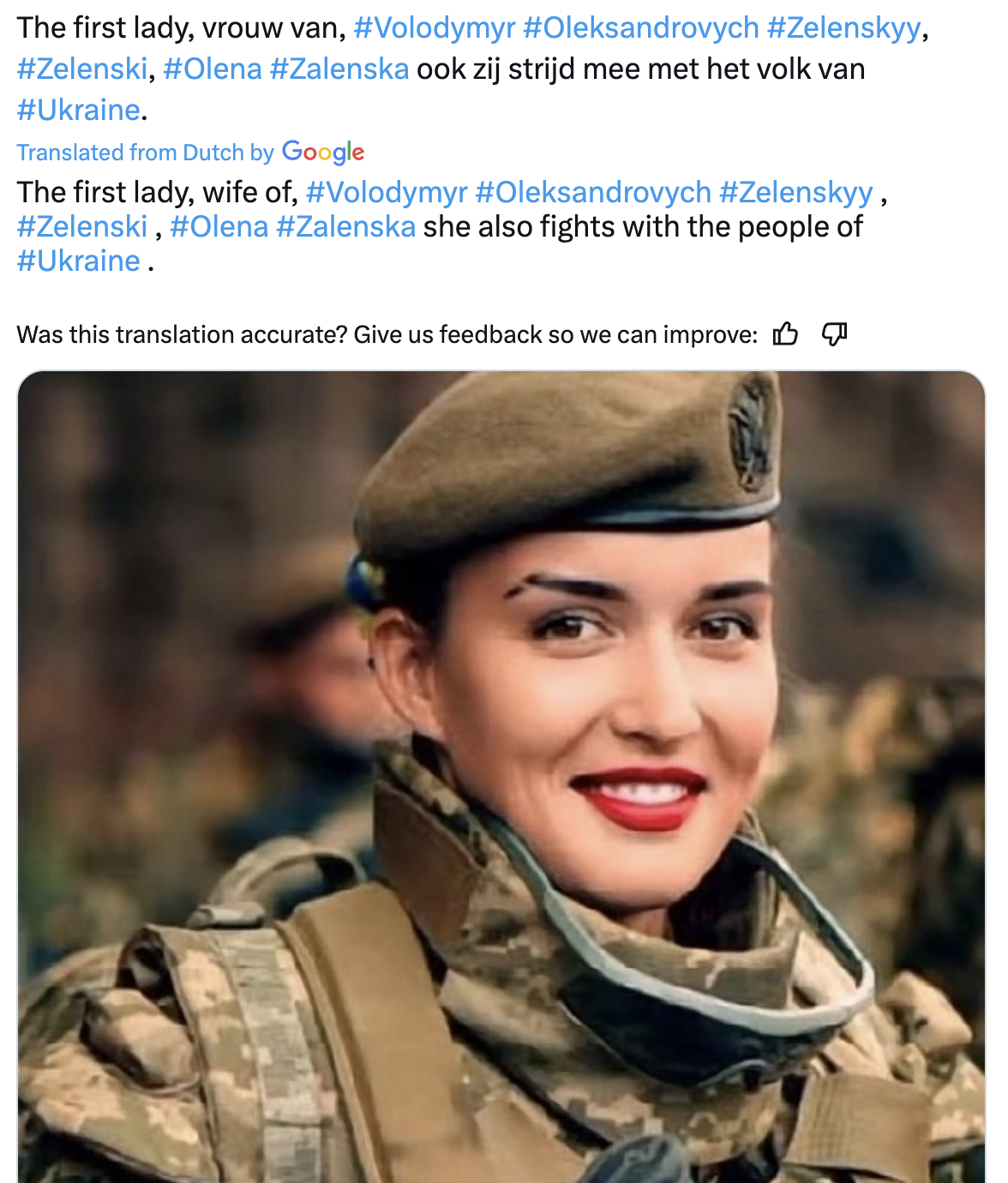}}
    \subfloat[Pro-Russia]{\includegraphics[width=0.24\textwidth]{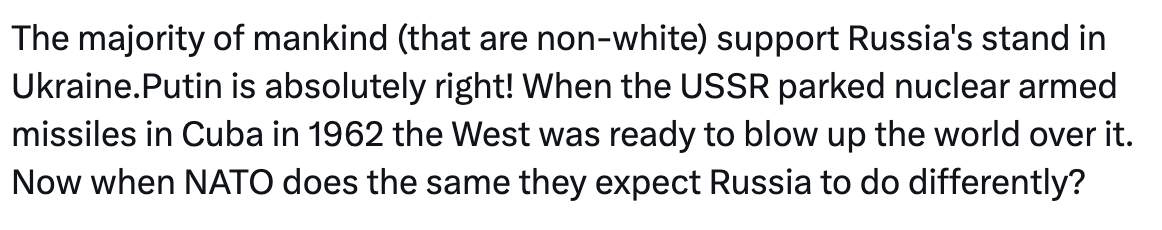}}\\
    
    \subfloat[Anti-Ukraine]{\includegraphics[width=0.24\textwidth]{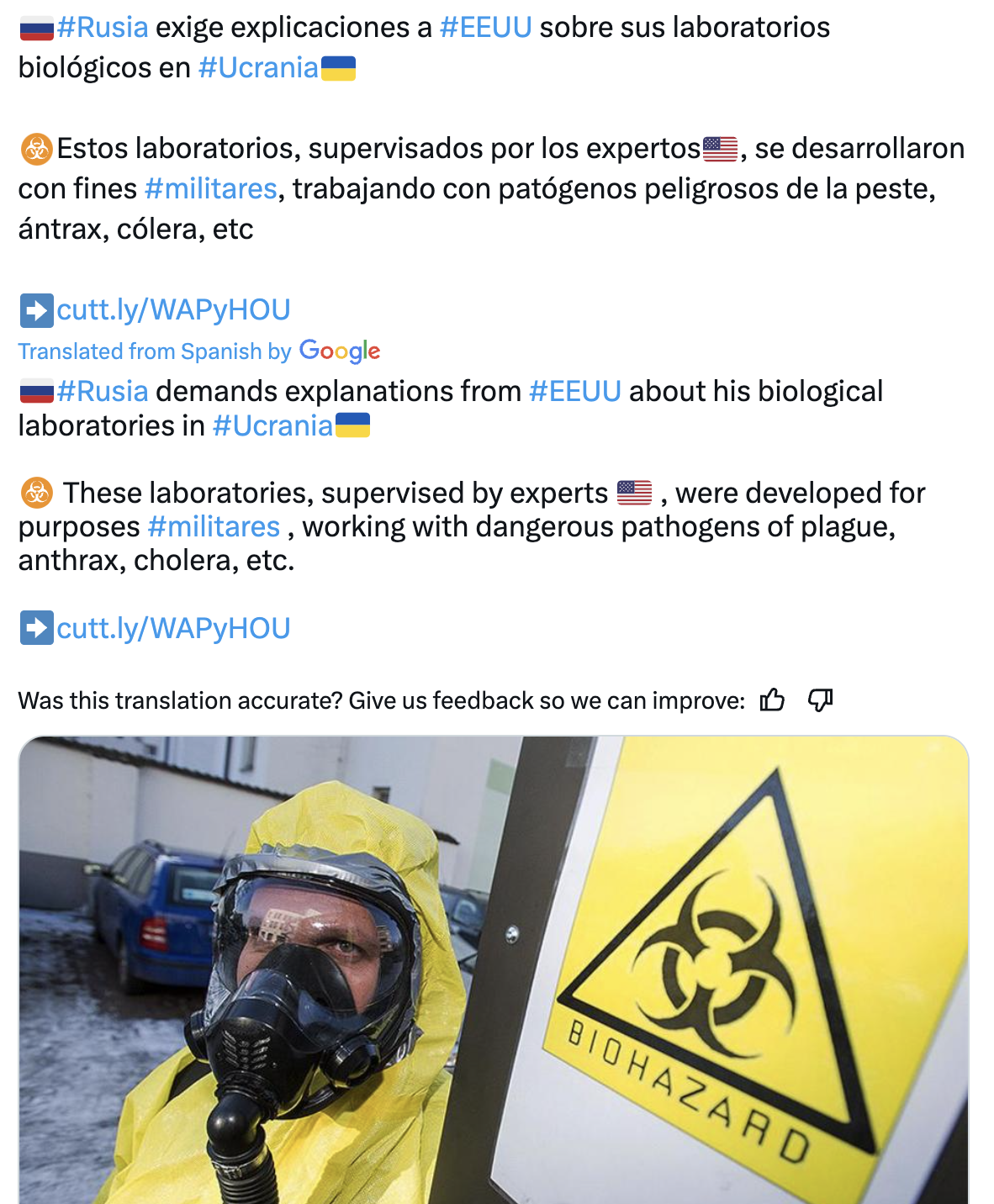}}
    \subfloat[Anti-Russia]{\includegraphics[width=0.24\textwidth]{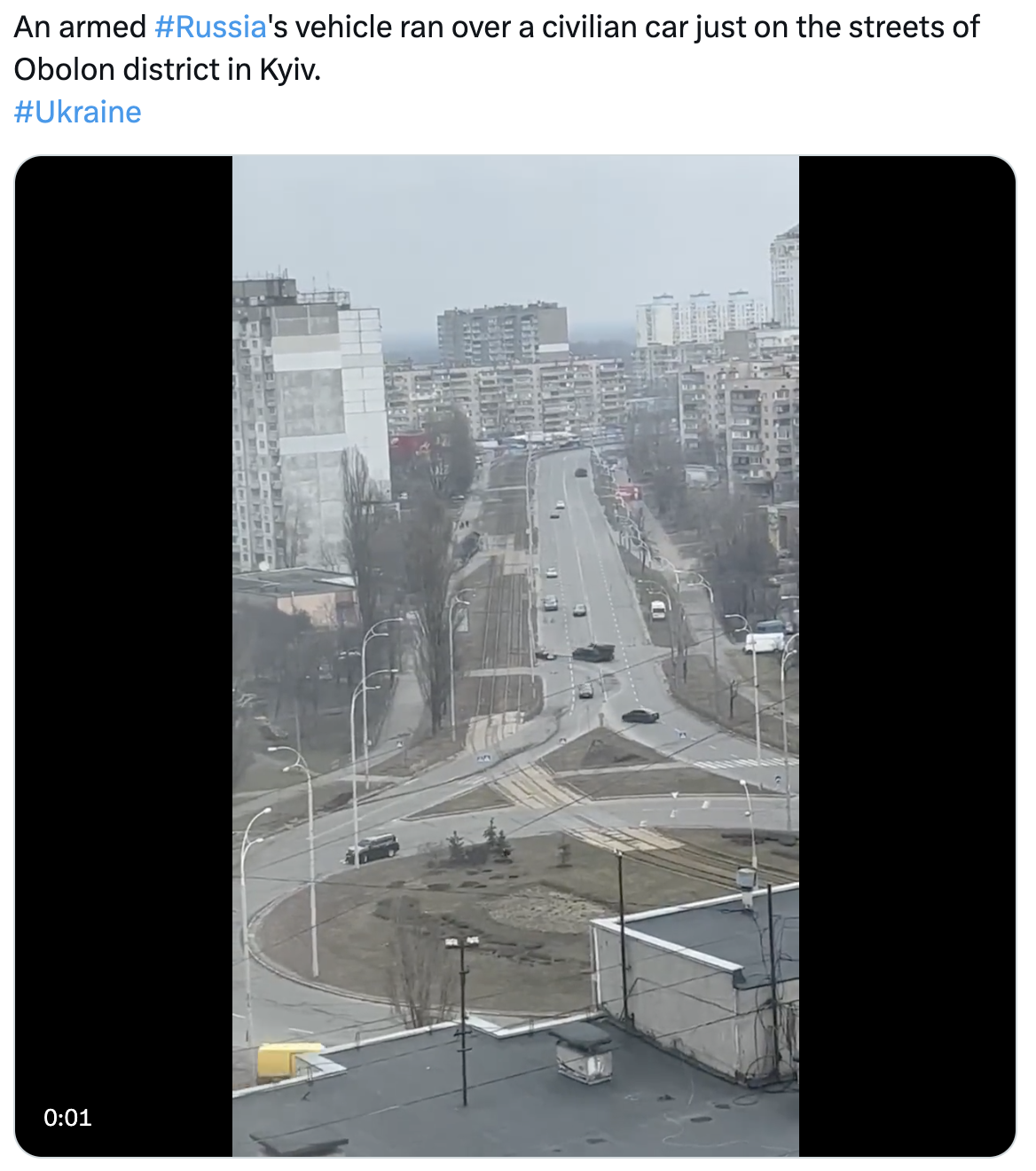}}
    \caption{Examples of tweets having specific narrative stances.}
    \label{fig:examples}
\end{figure}

\section{Keywords for Fact-checked Article Search}
\label{a:1}

The list of search keywords extracted from the Wikinews categories are as follows:
\emph{russian invasion of ukraine, russian invasion, invasion of ukraine, russo-ukrainian war, ukrainian war, controversies, international relations, russia, ukraine, europe, military history, military, military history of russia, history of russia, conflicts, invasions by russia, invasions russia, invasions, invasions of ukraine, opposition to nato, nato, conflicts in territory of the former soviet union, soviet union, former soviet union, russian irredentism, irredentism, russian–ukrainian wars, belarus–nato relations, belarus, belarus–ukraine relations, russia–nato relations, ukraine–nato relations, russia–nato, russia nato, ukraine nato, ukraine–nato, vladimir putin, putin, vladimir, volodymyr zelenskyy, volodymyr, zelenskyy, alexander lukashenko, alexander, lukashenko, wars involving belarus, wars involving Chechnya, wars involving russia, wars involving the donetsk people republic, wars involving the luhansk people republic, wars involving ukraine, donetsk, luhansk
}

\section{Keywords for Streaming Tweet Collection}
\label{a:2}
\begin{figure}[!htbp]
    \centering
    \includegraphics[width=1.01\linewidth]{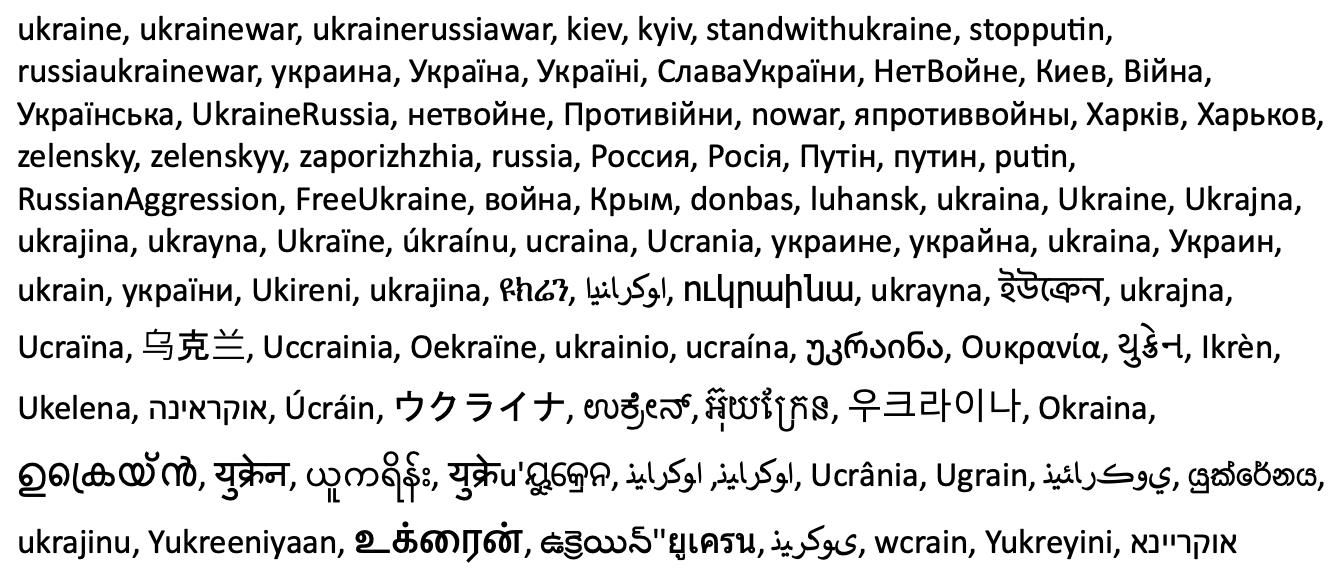}
    \caption{Keywords used for Streaming data collection.}
    \label{figure:keywords}
\end{figure}


\begin{table}[!htbp] 
  \caption{Examples of content translation for filtering fact-checked articles.}
  
  \begin{tabular}{p{3.5cm} p{1cm} p{3.5cm}}
    \toprule
    Original Title & Language & Translated Title \\
    \midrule
    Koncert w Odessie to dow\'od, \.zewojny w Ukrainie nie ma? & Polish &  The concert in Odessa is proof that there is no war in Ukraine?\\
    Selenskyj forderte die USA nicht auf, Soldaten in die Ukraine zu chicken & German & Zelensky did not ask the US to send soldiers to Ukraine \\

    Non, cette vid\'eo ne montre pas le "sosie cach\'e" du président ukrainien Zelensky, mais son garde du corps & French & No, this video does not show the "hidden double" of Ukrainian President Zelensky, but his bodyguard \\

    Forbes non ha scritto che gli aiuti finanziari all’Ucraina hanno arricchito i politici ucraini & Italian & Forbes did not write that financial aid to Ukraine enriched Ukrainian politicians \\


  \bottomrule
\end{tabular}
\label{tab:examples}
\end{table}

\end{document}